\documentclass[journal]{IEEEtran}
\usepackage{xcolor}
\usepackage{xurl}
\usepackage{soul}
\usepackage{booktabs}
\usepackage{tabularx}
\usepackage{graphicx}
\usepackage{makecell}
\usepackage{amsmath}
\usepackage{amssymb}
\usepackage{booktabs}
\usepackage{algorithm}
\usepackage{algorithmic}
\usepackage{xcolor}
\usepackage{hyperref}
\usepackage{tikz}
\usepackage{pgfplots}
\usepackage{multirow}
\usepackage{soul}
\usepackage{graphicx}
\usepackage[utf8]{inputenc}
\usepackage{textcomp}
\usepackage{amssymb}
\usepackage{pifont}
\usepackage{threeparttable}
\DeclareUnicodeCharacter{2248}{\ensuremath{\approx}} 
\DeclareUnicodeCharacter{2013}{--} 
\DeclareUnicodeCharacter{2014}{---} 
\usepackage{subcaption}
\usepackage{array}
\usepackage{adjustbox}
\usepackage{placeins}
\usepackage{float}
\usepackage[normalem]{ulem}
\usepackage[numbers]{natbib}
\renewcommand{\arraystretch}{0.9}
\definecolor{ucdblue}{RGB}{0, 47, 108}
\definecolor{ucdgold}{RGB}{255, 191, 0}

\hypersetup{
    colorlinks=true,
    linkcolor=ucdblue,
    citecolor=ucdblue,
    urlcolor=ucdblue
}
\pgfplotsset{compat=1.18}
\begin{document}

\title{Fleet Size and Mix Capacitated Vehicle Routing Problem with Time Windows for Mobile Fast Chargers

}

\author{Farhang~Motallebi Araghi,
        Armin~Abdolmohammadi,
        Navid~Mojahed,
        and~Shima~Nazari,~\IEEEmembership{Member,~IEEE}%
\thanks{The authors are with the Department of Mechanical and Aerospace Engineering, University of California at Davis, Davis, USA.}%
\thanks{F. Motallebi Araghi: \texttt{fmotallebi@ucdavis.edu}; A. Abdolmohammadi: \texttt{abdolmohammadi@ucdavis.edu}; N. Mojahed: \texttt{nmojahed@ucdavis.edu}; S. Nazari: \texttt{snazari@ucdavis.edu}.}%
\thanks{Manuscript received xxx xx, 2025.}}

\markboth{}%
{Motallebiaraghi \MakeLowercase{\textit{et al.}}: Fleet Size and Mix Capacitated Vehicle Routing Problem with Time Windows for Mobile Fast Charging Vehicles}

\maketitle 
\begin{abstract}
The electrification of off-road heavy equipment presents operational challenges for agencies serving remote sites with limited fixed charging infrastructure. Existing mobile fast charging vehicle (MFCV) planning approaches typically treat fleet design and routing as separate problems, fixing vehicle characteristics before dispatch. This paper formulates a fleet size and mix capacitated vehicle routing problem with time windows (FSMCVRPTW) for MFCV deployment, jointly optimizing fleet composition, charger specifications, routing, and scheduling within a unified mixed-integer linear program. The model incorporates heterogeneous MFCV types with varying power ratings, battery capacities, fuel range, and cost structures, minimizing total daily cost from labor, fuel, amortized capital expenditure, and energy purchase under temporal service windows, resource budgets, and energy-delivery constraints. The formulation is implemented in Python/Gurobi and applied to two case studies using California Department of Transportation wheel-loader data in Los Angeles (dense urban) and Truckee (sparse mountainous). Results show that simultaneous optimization yields compact, well-utilized fleets that meet all service windows while revealing strong sensitivity of unit cost to demand density and geography. The proposed FSMCVRPTW framework provides a generalizable decision-support methodology that co-designs fleet size, charger power, routing, and service schedules in a single optimization layer for context-aware, cost-efficient mobile fast charging.
\end{abstract}


\IEEEpeerreviewmaketitle

\begin{table}[!htbp]
\scriptsize
\centering
\caption{Mathematical Notation and Symbols}
\label{tab:notation_consolidated}
\scriptsize
\begin{tabularx}{\columnwidth}{@{}lX@{}}
\toprule
\textbf{Symbol} & \textbf{Definition} \\
\midrule
\multicolumn{2}{@{}l}{\textbf{Sets and Indices}} \\
$\mathcal{N}$ & Nodes (depot and client locations) \\
$\mathcal{J}$ & Clients requiring service \\
$\mathcal{C}$ & Clients used in waiting-time cost  \\
$\mathcal{T}$ & DCFC vehicle types \\
$\mathcal{K}$ & Universal set of vehicle slots across types \\
$i, j$ & Node indices \\
$\tau$ & Vehicle type index \\
$k$ & Vehicle slot index \\
$v$ & Vehicle index within type $\tau$ \\
\midrule
\multicolumn{2}{@{}l}{\textbf{Parameters}} \\
$d_{ij}$ & Distance between nodes $i$ and $j$ (miles) \\
$t_{ij}$ & Travel time between nodes $i$ and $j$ (hours) \\
${\mathcal{V}}$ & Nominal travel speed (miles/hour) \\
$E_j$ & Energy demand at client $j$ (kWh) \\
$B_j$ & Battery capacity of equipment at client $j$ (kWh) \\
$\rho_j$ & Power rating at client $j$ (kW) \\
$[a^{\min}_j, a^{\max}_j]$ & Off-hour service window for client $j$ \\
$P^{\max}_\tau$ & Max DC charging power for type $\tau$ (kW) \\
$B_\tau$ & Battery capacity for type $\tau$ (kWh) \\
$s_{j\tau}$ & Service duration at client $j$ by vehicle type $\tau$ (hours) \\
$F_\tau$ & Fuel capacity for type $\tau$ (fuel units) \\
$\phi_\tau$ & Fuel consumption rate (fuel units/mile) \\
$C^{\text{cap}}_\tau$ & Daily amortized capital cost per vehicle of type $\tau$ (USD/day) \\
$C_{\text{vehicle}}$ & Vehicle / chassis purchase cost in CAPEX decomposition (USD) \\
$C_{\text{DCFC}}$ & DC fast charger hardware cost (USD) \\
$C_{\text{trailer}}$ & Trailer or body / upfit cost (USD) \\
$C_{\text{battery}}$ & Battery pack cost used in CAPEX decomposition (USD) \\
$\text{Lifespan}_{\text{veh}}$ & Economic lifespan of vehicle / chassis / trailer (years) \\
$\text{Lifespan}_{\text{batt}}$ & Economic lifespan of battery pack (years) \\
$C^{\text{op}}_\tau$ & Operational cost per hour for type $\tau$ (USD/hour) \\
$\bar{v}_\tau$ & Max number of vehicles available of type $\tau$ \\
$v_\tau$ & Min required number of vehicles of type $\tau$ \\
$V^{\text{tot}}$ & Max total number of vehicles across all types \\
$t_{\text{start}}$ & Start time of planning horizon (hours) \\
$\sigma_{\text{batt}}, \sigma_{\text{fuel}}$ & Battery and fuel safety factors \\
$\alpha, \beta, \delta, \varepsilon, \zeta, \lambda_w$ & Objective-function weights \\
$M$ & Big-M parameter for logical constraints \\
\midrule
\multicolumn{2}{@{}l}{\textbf{Decision Variables}} \\
$y_{\tau v}$ & Binary: 1 if vehicle slot $v$ of type $\tau$ is selected \\
$x_{ijk}$ & Binary: 1 if vehicle $k$ travels from node $i$ to $j$ \\
$A_{ik}$ & Arrival time of vehicle $k$ at node $i$ (hours from $t_{\text{start}}$) \\
$W_i$ & Waiting time at node $i$ before service (hours; nonzero only for client nodes) \\
$L_j$ & Lateness slack at client $j$ (soft violation) \\
$u_{ik}$ & Visit order of node $i$ on route of vehicle $k$ (MTZ) \\

\bottomrule
\end{tabularx}
\end{table}

\section{Introduction}
\IEEEPARstart{T}{he} transportation sector’s shift to battery electric vehicles (BEVs) offers environmental benefits but introduces operational challenges for fleet operators, particularly in off-road applications \cite{IEA:2023, IPCC:2022}. For government agencies operating in diverse environments, the transition to electric heavy equipment requires innovative charging solutions.

As BEV technology advances, range anxiety remains a major barrier to large-scale adoption. In off-road heavy equipment operations, particularly at remote sites with limited or no charging infrastructure, this concern is amplified \cite{Nicholas:2021}, as downtime directly affects public infrastructure maintenance \cite{Franke:2013}. The consequences of such downtime extend beyond operational inconvenience, potentially compromising public safety (e.g., delayed emergency repairs or snow removal), delaying essential maintenance, and undermining confidence in electrification initiatives \cite{Dong:2014}.

Conventional approaches that require transporting depleted equipment to distant charging facilities lead to extended downtime and logistical challenges. In contrast, mobile fast charging vehicles (MFCVs) offer a promising alternative by providing high-power charging directly at remote operational sites, forming a charging network tailored to operational needs \cite{Wang:2020}. Efficient deployment of MFCVs involves complex trade-offs that require advanced optimization. Fleet managers must balance charger battery capacity, power rating, service area, and time-sensitive demand \cite{Erdogan:2012, Schneider:2014}, while also ensuring that MFCVs have sufficient energy by scheduling their own recharge intervals. The rest of the paper is organized as follows. Section \ref{sec:Literature} reviews the related work; Section \ref{sec:Problem_formulation} describes the problem formulation; Section \ref{sec:Solution Approach}  presents the solution approach and case study; Section \ref{sec:Computational results}  reports computational results; and Section \ref{sec: conslusions} concludes with insights and future work.

\section{Literature Review}
\label{sec:Literature}
The operational challenges of mobile fast charging fall within the family of vehicle routing and scheduling problems, which have been widely studied in operations research. The vehicle routing problem (VRP), a foundational model for fleet routing, was introduced by Dantzig and Ramser (1959) as the “Truck Dispatching Problem”~\cite{dantzig1959truck}. It generalizes the traveling salesman problem (TSP) by using a fleet of vehicles instead of a single salesman and seeks optimal routes from one or more depots to serve multiple customers while minimizing distance or cost~\cite{laporte2013vehicle}. As the TSP is NP-hard, the VRP inherits this complexity, motivating continued algorithmic advances.

The classical VRP has been extended to incorporate practical constraints such as capacity, time windows, and energy limits. The capacitated VRP (CVRP) restricts how much each vehicle can serve per route and is relevant in industries with strict weight or volume limits. The VRP with time windows (VRPTW) requires servicing customers within predefined time intervals and became widely adopted after Solomon’s 1987 benchmark datasets, which remain standard for evaluating routing algorithms~\cite{solomon1987algorithms}. VRPTW jointly optimizes routes and service times: vehicles may wait if they arrive before a customer’s time window or time arrivals to meet just-in-time delivery requirements, making it central for logistics operations with strict service commitments.

The multi-depot VRP (MDVRP) extends this framework to scenarios with multiple depots. Early work includes exact methods by Laporte and Arpin (1984) and heuristics by Tillman (1969)~\cite{laporte1984optimal, tillman1969multiple}. MDVRP increases complexity by coupling customer–depot assignment with routing, and is particularly relevant for organizations operating multiple service hubs. While numerous other variants exist (e.g., pickup-and-delivery, open routing, stochastic demands), CVRP and VRPTW remain core building blocks for advanced fleet optimization models~\cite{desaulniers2002vrp, bertsimas1992vehicle}. Fleet optimization also includes strategic decisions such as fleet sizing and facility location. Fleet size and mix problems determine the optimal number and types of vehicles by balancing fixed fleet costs with routing costs~\cite{toth2014vehicle}. Golden et al. (1984) extended the VRP to heterogeneous fleets, enabling joint vehicle selection and routing~\cite{golden1984fleet}. Facility location models, in turn, optimize depot or service center placement and customer allocation to minimize cost and distance~\cite{arabani2012facility}.

The electric vehicle routing problem (EVRP) extends VRP to account for limited battery capacity and en-route recharging~\cite{kucukoglu2021electric}. Erdogan and Miller-Hooks (2012) introduced the Green VRP, allowing refueling at alternative stations~\cite{erdougan2012green} and laying the foundation for routing with energy constraints. Schneider et al. (2014) incorporated time windows and recharging stations (E-VRPTW), jointly modeling routing, battery state-of-charge, and service time requirements~\cite{schneider2014electric}. Subsequent studies added features such as partial recharging and multiple charging technologies~\cite{felipe2014heuristic}, explicit battery behavior and degradation modeling~\cite{montoya2017electric}, and refined recharging strategies under time pressure~\cite{keskin2016partial}. Mixed-fleet settings with both electric and conventional vehicles have also been studied~\cite{goeke2015routing, hiermann2016electric}. Across these extensions, the central challenge is balancing route efficiency with battery limitations and charging logistics.

Most EVRP studies assume fixed charging infrastructure, but recent research has investigated mobile charging solutions that dynamically respond to demand. Initial work considered emergency roadside support services~\cite{afshar2021mobile}. Huang et al. (2014) proposed mobile charging via plug-in or battery-swapping vehicles that travel to EVs in need, developed a queuing-based analytical framework, and evaluated dispatch strategies such as the nearest-job-next rule~\cite{huang2014design}. Subsequent studies examined design parameters, identifying suitable ranges for onboard energy capacity and charging power~\cite{ruaboacua2020optimization}.

More recent work treats mobile charging stations as structured, scalable services. Cui et al. formulated the mobile charging VRP with time windows (MC-VRPTW), extending CVRPTW with energy transfer constraints between mobile chargers and BEVs~\cite{cui2018mobile}. Afshar et al. integrated fixed and mobile chargers in urban settings, showing reduced wait times and grid stress under a fixed mobile fleet configuration~\cite{afshar2022optimal}. Tang et al. proposed a two-layer framework that jointly optimizes depot locations, fleet size, battery capacity, and routing under stochastic demand~\cite{tang2020online}, while Beyazıt and Taşcıkaraoğlu developed a city-scale mixed-integer linear programming (MILP) model incorporating electricity pricing and battery degradation~\cite{beyazit2023electric}. These studies demonstrate the potential of coordinated mobile charging strategies but typically rely on predetermined fleet designs or face economic limitations.

Advances in artificial intelligence have further shifted mobile charging optimization from classical routing to autonomous dispatch strategies. Related multi-agent work under incomplete information shows that coordination can depend strongly on limited or evolving system knowledge \cite{Yousefgomgashte}, supporting the move toward decentralized decision-making. Liu et al. modeled mobile charging station scheduling as a decentralized multi-agent system, where spatial-aware multi-agent deep reinforcement learning enables MCVs and BEVs to make local real-time decisions, improving charging success rates and operator revenue over heuristic baselines~\cite{liu2024multi}. Mobile charging has also been embedded into infrastructure planning. Xu et al. developed a dynamic expansion model that integrates fixed chargers with rental-based MCS deployment under stochastic demand using sample average approximation, improving charger utilization and reducing investment risk~\cite{xu2025dynamic}. Duan et al. incorporated MCS operations into an integrated energy system that coordinates EV charging with photovoltaic generation, grid support, and energy storage; their MILP-based multi-scenario model reduced daily operating costs by over 90\% and significantly decreased renewable energy curtailment~\cite{duan2025study}.

\subsection{Research Gaps and Contributions}

As summarized in Table \ref{tab:literature_comparison_updated}, existing studies typically separate infrastructure planning (e.g., fleet size and battery capacity) from operational decisions such as routing and scheduling.  Such frameworks do not perform simultaneous optimization across all relevant decision variables, limiting their ability to capture interdependencies between charger design and operational strategy. Other studies assume fixed fleet characteristics or ignore vehicle heterogeneity, which further limits their applicability. Only a small number of works simultaneously consider fleet sizing, routing, and scheduling, and even these do not solve the problem within a single integrated optimization model. For example, Tang et al. (2020) \cite{tang2020online} adopt a bi-level framework in which fleet configuration is determined prior to dispatch, leading to sequential rather than fully joint decision-making. To the best of our knowledge, no existing framework jointly determines the number of mobile charging vehicles, their technical specifications, routing paths, and scheduling strategies within a single, unified optimization formulation.

This paper addresses this limitation through a joint design–operation optimization framework that integrates MFCV sizing and routing within an extended multi-objective CVRPTW formulation. Using real heavy-equipment operational data, we evaluated the economic feasibility of deployment under urban and rural conditions.

\newcommand{\xmark}{\ding{55}}
\renewcommand{\arraystretch}{0.9}
\begin{table}[t]
\centering
\caption{Literature classification: Fleet Sizing, Routing, and Scheduling}
\label{tab:literature_comparison_updated}
\scriptsize
\setlength{\tabcolsep}{3pt}
\renewcommand{\arraystretch}{0.85}
\begin{tabularx}{\columnwidth}{@{}lccc X@{}}
\toprule
\textbf{Reference} & \textbf{Size} & \textbf{Route} & \textbf{Sched.} & \textbf{Comment} \\
\midrule
\multicolumn{5}{@{}l}{\textit{Scheduling optimization only}} \\
\midrule
\cite{cui2018mobile}         &        & \checkmark & \checkmark & Fixed MFCV fleet \\
\cite{pelletier2019electric} &        & \checkmark & \checkmark & No fleet sizing \\
\cite{schneider2014electric} &        & \checkmark & \checkmark & Fixed stations \\
\cite{zhang2020fuzzy}        &        & \checkmark & \checkmark & Fixed parameters \\
\cite{afshar2022optimal}     &        & \checkmark & \checkmark & No design vars \\
\cite{verma2018electric}     &        & \checkmark & \checkmark & Predefined specs \\
\cite{basso2019energy}       &        & \checkmark & \checkmark & Fixed fleet/setup \\
\midrule
\multicolumn{5}{@{}l}{\textit{Fleet sizing / design only}} \\
\midrule
\cite{schiffer2017electric}  & \checkmark & \checkmark &           & Sequential design/ops \\
\cite{Schneider2018-zp}      & \checkmark &           &           & Decoupled fleet–route \\
\cite{fisher1981generalized} & \checkmark & \checkmark &          & No TW / energy \\
\cite{salhi2013fleet}        & \checkmark & \checkmark &          & limited design–sched \\
\midrule
\multicolumn{5}{@{}l}{\textit{Partial integration}} \\
\midrule
\cite{tang2020online}        & \checkmark & \checkmark & \checkmark & Bi-level, sequential \\
\cite{zhang2025joint}        & \checkmark & \checkmark & \checkmark & Fixed infra, simplified \\
\cite{yang2023fleet}         & \checkmark & \checkmark & \checkmark & Limited design space \\
\midrule
\textbf{This work}           & \textbf{\checkmark} & \textbf{\checkmark} & \textbf{\checkmark} &
\textbf{Single-level co-optimization of MFCV sizing, routing, scheduling} \\
\bottomrule
\end{tabularx}
\end{table}

\section{Problem Formulation}
\label{sec:Problem_formulation}
This section presents the mathematical formulation of the optimization model. It begins by outlining the underlying assumptions of the study, which define the modeling framework. The formulation then specifies the network structure, client characteristics, vehicle specifications, decision variables, objective function, and full constraint system.

\subsection{System Setting and Assumptions}

\subsubsection{\textbf{Model Assumptions}}

The mathematical formulation is based on the following assumptions: deterministic travel times and distances; full vehicle reliability over the planning horizon; fixed client energy requirements; rigid service windows; constant charging efficiency; complete information availability; single-day operational cycles; single-client service per vehicle per visit; uniform fuel consumption rates within each vehicle type; and deterministic service durations.

\subsubsection{\textbf{Network and Index Sets}}

The service network is defined on the node set $\mathcal{N} = \{0\} \cup \mathcal{J}$, where node $0$ is the central depot for all MFCVs and $\mathcal{J} = \{1,2,\ldots,n\}$ denotes client locations requiring off-hour charging. Vehicle types are collected in $\mathcal{T} = \{\textbf{Standard, Medium, High, Ultra, Mega}\}$ with indices $\{1,2,3,4,5\}$; each type $\tau$ has distinct power, battery, fuel, and cost parameters summarized in Table~\ref{tab:mobile_dcfc_specs_corrected}. Practical availability imposes upper bounds $\bar{v}_\tau \in \{10,10,8,5,3\}$ for $\tau \in \{1,2,3,4,5\}$. The universal vehicle-slot set $\mathcal{K}$ contains all potential vehicles, with $|\mathcal{K}| = \sum_{\tau \in \mathcal{T}} \bar{v}_\tau = 36$. Each slot $k \in \mathcal{K}$ corresponds to a unique vehicle of type $\tau(k) \in \mathcal{T}$ and index $v(k) \in \{1,\dots,\bar{v}_{\tau(k)}\}$, forming a one-to-one pair $(\tau(k), v(k))$. Geographic inputs include distance $d_{ij}$ (miles) between nodes $i$ and $j$ and travel time $t_{ij}$ (hours) required to traverse arc $(i,j)$ at the operational vehicle speed.

\begin{table}[!t]
\centering
\caption{MFCV Specifications, Cost Structure, and Objective Coefficients}
\label{tab:mobile_dcfc_specs_corrected}
\resizebox{\columnwidth}{!}{%
\renewcommand{\arraystretch}{0.85}
\begin{tabular}{@{}lrrrrr@{}}
\toprule
\textbf{Specification} & \textbf{Std.} & \textbf{Med.} & \textbf{High} & \textbf{Ultra} & \textbf{Mega} \\ 
\midrule
\multicolumn{6}{@{}l}{\textit{Mobile Battery Technical Specifications}} \\
$P^{\max}_\tau$ (kW)           & 50   & 200  & 350  & 500  & 1{,}000 \\
$B_\tau$ (kWh)                 & 80   & 160  & 300  & 500  & 1{,}000 \\
\midrule
\multicolumn{6}{@{}l}{\textit{Vehicle and Trailer Capital Cost and Specifications}} \\
$C_\text{Trailer} + C_\text{Vehicle}$ (\$K)  & 80  & 80  & 80  & 80  & 80 \\ 
$C_\text{DCFC}$ (\$K)                        & 100 & 250 & 450 & 650 & 1{,}200 \\
$F_\tau$ (gal)                               & 40  & 60  & 80  & 100 & 150 \\
$\phi_\tau$ (gal/mile)                         & 0.10& 0.12& 0.15& 0.18& 0.25 \\
\midrule
\multicolumn{6}{@{}l}{\textit{Daily Amortized CAPEX (\$/day)}} \\
$C^{\text{cap}}_\tau$                       & 65.75 & 147.95 & 258.64 & 367.12 & 668.59 \\
\midrule
\multicolumn{6}{@{}l}{\textit{Operational Costs}} \\
$C^{\text{op}}_{\tau}$ (\$/hr)              & 1.0 & 1.2 & 1.5 & 1.8 & 2.5 \\
\midrule
\multicolumn{6}{@{}l}{\textit{Objective Coefficients}} \\
\multicolumn{6}{@{}p{\linewidth}}{\raggedright
$\alpha=30$ (\$/h), $\lambda_w=30$ (\$/h), $\beta=3.80$ (\$/gal), $\delta=100$ (\$/h), $\varepsilon=1.0$, $\zeta=1.0$, $\gamma=0.10$ (\$/kWh), $D_\text{year} = 365$.
} \\
\midrule
\multicolumn{6}{@{}l}{\textit{Efficiency Coefficients (battery and fuel)}} \\
\multicolumn{6}{@{}p{\linewidth}}{\raggedright
$\sigma_{\text{batt}} = 0.9$, $\sigma_{\text{fuel}} = 0.9$
} \\
\bottomrule
\end{tabular}%
}
\end{table}

\subsubsection{\textbf{Client Requirements and Service Windows}}

Each client $j \in \mathcal{J}$ is characterized by operational and energy
requirements that determine the MFCV service parameters. The service window
$[a^{\min}_j, a^{\max}_j]$ (hours from midnight) denotes the off-hour period
during which client $j$ can be served. These windows are based on observed schedules and include safety buffers to ensure charging finishes before operations resume.

$E_j$ (kWh) is the required energy delivered to client $j$ and is treated as a fixed input based on operational records. $\rho_j$ (kW) denotes the maximum charging power that client $j$ can accept, as determined by the on-board charging system, thereby setting an upper bound on the power transfer rate.

Any MFCV type can serve any client. For a vehicle of type $\tau$, the effective DC charging power at client $j$ is
\begin{equation}
p_{j\tau} = \min\big(P^{\max}_\tau, \rho_j\big),
\end{equation}
where $P^{\max}_\tau$ is the charger-side power limit (kW) for type $\tau$.
Given required energy $E_j$, the corresponding deterministic service time is
\begin{equation}
s_{j\tau} = \frac{E_j}{p_{j\tau}}
          = \frac{E_j}{\min\big(P^{\max}_\tau, \rho_j\big)}.
\end{equation}
Two scenarios arise: in the charger-limited case ($P^{\max}_\tau \le \rho_j$), increasing $P^{\max}_\tau$ reduces dwell time; in the client-limited case ($P^{\max}_\tau > \rho_j$), additional charger power offers no benefit. Thus, $P^{\max}_\tau$ affects routing feasibility via $s_{j\tau}$, allowing more clients per tour but at a higher capital cost $C^{\text{cap}}_\tau$.

\subsubsection{\textbf{Mobile Charger and Vehicle  Specifications}}
Each DCFC class is defined by its charging performance and energy capacity, as shown in Table~\ref{tab:mobile_dcfc_specs_corrected}. The maximum DC charging power $P^{\max}_\tau$ (kW) determines the achievable charging rate, while the on-board battery capacity $B_\tau$ (kWh) sets the total energy available for client service per route. Each DC fast charger is then mounted on a truck–trailer configuration with fuel capacity of $F_\tau$ (gal) and consumption rate of $\phi_\tau$ (gal/mile), which determine the operational travel range.

\subsubsection{\textbf{Decision Variables}}
The model employs five categories of decision variables. Fleet selection is represented by $y_{\tau v} \in \{0,1\}$, indicating whether slot $v$ of vehicle type $\tau$ is activated. Routing decisions use $x_{ijk} \in \{0,1\}$ to denote whether vehicle $k$ travels from node $i$ to node $j$. Temporal variables $A_{ik} \in [0,24]$ capture arrival times (hours since $t_{\text{start}}$). Service quality is modeled using waiting time $W_i \ge 0$ and lateness $L_j \ge 0$. 

Subtour-elimination following the Miller–Tucker–Zemlin (MTZ)~\cite{Desrochers:1991} formulation, introduces an additional set of integer decision variables $u_{jk}$, defined as

\begin{equation}
u_{jk} \in \{1,2,\ldots,|\mathcal{J}|\} \qquad \forall j \in \mathcal{J},\; \forall k \in \mathcal{K}.
\label{eq:mtz_domain}
\end{equation}

These variables are only defined for client nodes, since the depot implicitly acts as route origin and terminus. Their role in preventing subtours is discussed in Subsection~\ref{susubsection: Subtour Elimination}.

\subsection{Cost Function Structure and Coefficients}

Here, we formulate the objective function, which minimizes the total daily cost of operating the mobile charging fleet. The optimization objective is expressed as a weighted sum of six cost components, all modeled on a daily basis. These components include travel and service labor, fuel consumption, waiting and lateness penalties, amortized capital investment, and time-based operational expenses. The corresponding coefficients and parameter values are provided in Table~\ref{tab:mobile_dcfc_specs_corrected}.

In addition to the optimization-driven cost components, the model also accounts for the energy transfer cost associated with electricity purchased for DC fast charging. Since the total delivered energy is fixed across all feasible solutions, this term is excluded from the cost function formulation but is utilized in the final economic assessment and reported in the results in subsection~\ref{Economic_Alanysis}.

\subsubsection{\textbf{Optimization Objective Function Components}}
The optimization objective minimizes a composite daily cost in USD expressed as a sum of six components:
\begin{equation}
\begin{aligned}
\min \text{TC} ={}& C_{\text{Travel and Service}} + C_{\text{fuel}} + C_{\text{wait}} \\
                  & + C_{\text{late}} + C_{\text{CAPEX}} + C_{\text{OPEX}}.
\end{aligned}
\end{equation}

The travel and service time cost captures operator labor during active driving and charging:
\begin{equation}
C_{\text{Travel and Service}}
  = \alpha \sum_{k} \sum_{i \neq j} (t_{ij} + s_{j\tau(k)}) \, x_{ijk},
\end{equation}
where $\alpha$ (USD/h) is the labor cost coefficient, $t_{ij} = d_{ij}/\mathcal{V}$
is the travel time between nodes $i$ and $j$, and $s_{j\tau(k)}$ is the
deterministic service time for client $j$ served by the type of vehicle $k$.
This term accounts for productive time spent moving and delivering energy.

Waiting time cost is modeled as
\begin{equation}
C_{\text{wait}} = \lambda_w \sum_{j \in \mathcal{J}} W_j,
\end{equation}
where $\lambda_w$ (USD/h) is the cost of operator idle time and $W_j$
denotes waiting time at client $j$. This penalty discourages early arrivals
and promotes schedules that align arrivals with service window openings.

Fuel cost is given by
\begin{equation}
C_{\text{fuel}} = \beta \sum_{k} \sum_{i \neq j} d_{ij} \,\phi_{\tau(k)} \, x_{ijk},
\end{equation}
where $\beta$ (USD/gal) is the diesel price and $\phi_{\tau}$ (gal/mile) is the
type-specific fuel consumption rate.
This term internalizes the energy cost of moving MFCVs between clients and the depot.

Lateness penalties are modeled as
\begin{equation}
C_{\text{late}} = \delta \sum_{j \in \mathcal{J}} L_j,
\end{equation}
where $\delta$ (USD/h) is a penalty weight and $L_j$ denotes lateness at
client $j$. This cost acts as a soft service-level constraint, making late
service economically unattractive while preserving feasibility when strict
time-window adherence is impossible.

The daily amortized capital expenditure (CAPEX) term is
\begin{equation}
C_{\text{CAPEX}} = \varepsilon \sum_{\tau \in \mathcal{T}} \sum_{v} C^{\text{cap}}_\tau \, y_{\tau v},
\end{equation}
where $\varepsilon$ is a scaling weight and $C^{\text{cap}}_\tau$ (USD/day)
is the daily amortized capital cost for vehicle type $\tau$, computed as
\begin{equation}\label{eq_C_cap}
C^{\text{cap}}_\tau =
\frac{1}{D_{\text{year}}}
\left[
\frac{C_{\text{vehicle}} + C_{\text{trailer}}}{\text{Lifespan}_{\text{veh}}}
+
\frac{C_{\text{DCFC}}}{\text{Lifespan}_{\text{DCFC}}}
\right],
\end{equation}
with $D_{\text{year}}$ denoting the number of operational days per year and
$\text{Lifespan}_{\text{veh}}$ and $\text{Lifespan}_{\text{DCFC}}$ denoting the
assumed economic lifetimes of the base vehicle/trailer and DC fast charging
equipment, respectively. This term captures the long-run
investment cost of deploying MFCVs.

Operational expenditure (OPEX) is
\begin{equation}
C_{\text{OPEX}} = \zeta \sum_{k} \sum_{i \neq j} C^{\text{op}}_{\tau(k)} \, t_{ij} \, x_{ijk},
\end{equation}
where $\zeta$ is a scaling weight and $C^{\text{op}}_{\tau}$ (USD/h) represents
type-specific operational costs (e.g., maintenance, insurance, and power
electronics wear) as summarized in
Table~\ref{tab:mobile_dcfc_specs_corrected}. This term accounts for
time-dependent operating expenses beyond fuel and labor.

\subsubsection{\textbf{Non-Optimizing Cost Component-Energy Transfer}}

The energy transfer term is
\begin{equation}
C_{\text{Energy\_transfer}}
  = \gamma \sum_{k \in \mathcal{K}} \sum_{i \neq j} E_j \, x_{ijk},
\end{equation}
where $\gamma$ (USD/kWh) represents the electricity purchase cost for DC fast
charging. Because the model assumes that all client energy demands are fully met, the total delivered energy is fixed across feasible solutions, so $C_{\text{Energy\_transfer}}$ does not influence the optimization decisions. It is therefore excluded from the optimization objective and added when reporting the  full daily economic costs in the results in subsection~\ref{Economic_Alanysis}.

\subsection{Optimization Constraints}
This section presents the complete system of constraints that govern routing, scheduling, fleet composition, and operational feasibility within the optimization model. The mathematical formulation incorporates eight major constraint categories:

\subsubsection{\textbf{Service Coverage}}
The universal service constraint ensures each client is visited exactly once:
\begin{equation}
\sum_{k \in K} \sum_{i \in N, i \neq j} x_{ijk} = 1 \quad \forall j \in J.
\end{equation}

\subsubsection{\textbf{Fleet Deployment}}
The fleet deployment constraints define the depot-based routing framework used for centralized MFCV operations \cite{Toth:2014}, enforcing that each selected vehicle must depart from and return to the central depot in accordance with the hub-and-spoke distribution model commonly applied in logistics systems \cite{Laporte:2009}. This requirement is captured by:
\begin{equation}
\sum_{j \in J} x_{0jk} = \sum_{i \in J} x_{i0k} = y_{\tau(k),v(k)} \quad \forall k \in \mathcal{K}.
\end{equation}

This constraint serves two purposes: (i) it links fleet composition to routing decisions by restricting route assignment to vehicles that are selected in the active fleet, and (ii) it preserves the depot-based routing structure by requiring every active vehicle to complete a full round trip \cite{Golden:2008}.

\subsubsection{\textbf{Network Flow Conservation}}
Network flow conservation constraints preserve route continuity throughout the transportation network, preventing vehicles from transitioning between non-adjacent nodes without traversing intermediate locations \cite{Ahuja:1993}. As a fundamental requirement in vehicle routing, this condition maintains physical realism by enforcing flow balance at each visited node.

This constraint is expressed as:
\begin{equation}
\sum_{i \in N, i \neq j} x_{ijk} = \sum_{i \in N, i \neq j} x_{jik} \quad \forall j \in J,\; k \in \mathcal{K}.
\end{equation}

Here, the total flow into client $j$ by vehicle $k$ must equal the total flow departing from $j$, thus enforcing classical flow conservation as established in network optimization theory \cite{Bertsekas:1998}.

\subsubsection{\textbf{Subtour Elimination}}
\label{susubsection: Subtour Elimination}
In vehicle routing problems, a \emph{subtour} refers to a closed tour that includes only a subset of client nodes without passing through the depot, leading to an infeasible solution. To prevent such disconnected cycles, the classical Miller–Tucker–Zemlin formulation is adopted \cite{Miller:1960,Desrochers:1991}. It enforces that each vehicle forms a single, continuous route beginning and ending at the depot. The MTZ constraints are:
\begin{equation}
u_{ik} - u_{jk} + M\, x_{ijk} \;\leq\; M - 1 
\qquad \forall k \in \mathcal{K},\; i \neq j \in \mathcal{J},
\end{equation}
where $u_{jk}$ are order indices over clients with standard bounds
\begin{equation}
u_{jk} \in \{1,2,\ldots,|\mathcal{J}|\}
\qquad \forall j \in \mathcal{J},\; k \in \mathcal{K}.
\end{equation}
The constant $M$ is a big-$M$ parameter that relaxes the ordering constraint when arc $(i,j)$ is not used ($x_{ijk}=0$) and enforces $u_{jk} > u_{ik}$ when the arc is active ($x_{ijk}=1$). A tight and numerically stable choice is $M = |\mathcal{J}|$, since the largest possible difference between order indices is bounded by the number of clients. Larger values may degrade the linear relaxation and lead to numerical instability \cite{Christofides:1981}.

\subsubsection{\textbf{Fleet Selection Coupling}}
Fleet selection coupling constraints link fleet deployment choices to routing feasibility, aligning strategic vehicle availability with tactical routing decisions \cite{Salhi:2014}. This integration is critical in FSMCVRPTW, where routing can only occur using actively deployed vehicles. Routing–fleet selection linkage is enforced by:
\begin{equation}
x_{ijk} \le y_{\tau(k), v(k)} \quad \forall i \ne j,\; k \in \mathcal{K}.
\end{equation}

This constraint guaranties that vehicle $k$ may only traverse arc $(i,j)$ if its corresponding slot has been selected for the active fleet, thereby synchronizing deployment and operational utilization \cite{Prins:2004}.

\subsubsection{\textbf{Fleet Size Management}}
Fleet size management constraints control the composition and scale of the MFCV fleet to accommodate practical operational limitations \cite{Clarke:1964,Fisher:1981}. These constraints can address business requirements, including budget constraints, depot capacity limitations, staffing restrictions, and regulatory limits on commercial vehicle operations. This set of constraints is given as:
\begin{align}
\sum_v y_{\tau v} &\geq v_\tau \quad \forall \tau \in \mathcal{T}, \label{eq:constraint_min} \\
\sum_{\tau,v} y_{\tau v} &\leq V^{\text{tot}}. \label{eq:constraint_max}
\end{align}

Constraint~\eqref{eq:constraint_min} enforces the minimum fleet composition requirement, whereas Constraint~\eqref{eq:constraint_max} limits the total number of deployed vehicles. 

\subsubsection{\textbf{Temporal and Service Window Constraints}}
Scheduling requirements are modeled through temporal constraints that regulate vehicle arrival times and enforce client service windows during off-hour periods \cite{Solomon:1987}. These constraints are essential for preventing operational conflicts with clients schedules. To establish a unified temporal reference, the arrival time at the depot is fixed at the start of operations:
\begin{equation}
A_{0,k} = t_{\text{start}} = 0,
\end{equation}

providing a consistent origin for all time-based calculations in the model \cite{Desaulniers:2005}. Time propagation along the route is governed by:

\begin{align}
A_{j,k} \geq A_{i,k} + W_i + s_{ik} 
&\; +\: t_{ij} - M(1 - x_{ijk}) \nonumber \\
&\qquad\forall\, i \neq j,\; k \in \mathcal{K}, 
\label{eq:time_propogation}
\end{align}

where \eqref{eq:time_propogation} ensures temporal consistency by requiring that if a vehicle $k$ travels from node $i$ to node $j$, it must complete all activities at the node $i$ (including waiting and service) plus travel time before arriving at node $j$ \cite{Cordeau:2007}.

Service window compliance is enforced through:
\begin{align}
A_{j,k} + W_j &\geq a^{\min}_j \quad \forall j \in \mathcal{J}, k \in \mathcal{K}, \label{eq:service_start}\\
A_{j,k} + W_j + s_{jk} &\leq a^{\max}_j + L_j + M\left(1 - \sum_i x_{ijk}\right) \nonumber \\
&\quad \forall j \in \mathcal{J}, k \in \mathcal{K}, \label{eq:service_end}
\end{align}

where \eqref{eq:service_start} prevents service from beginning before the allowable window opens, whereas \eqref{eq:service_end} permits controlled violations via slack variable $L_j$, thus maintaining feasibility while penalizing service outside the allowed time window \cite{Braysy:2005}.

\subsubsection{\textbf{Resource Constraints}}
Building on previous studies~\cite{Schneider:2014,Schiffer:2018,Erdogan:2012,Wang:2020}, the resource constraint formulation limits energy delivery and travel to 
respect battery and fuel capacities. The battery energy budget for each vehicle is given by:
\begin{align}
\sum_{j \in \mathcal{J}} 
E_j\!\left( \sum_{i \in \mathcal{N},\, i \neq j} x_{ijk} \right)
&\;\leq\;
B_{\tau(k)} \,\sigma_{\text{batt}} \, y_{\tau(k),v(k)}
\nonumber \\
&\qquad\forall\, k \in \mathcal{K},
\label{eq:battery_capacity}
\end{align}

where $\sigma_{\text{batt}}$ is the battery efficiency during the charging session. This inequality forces the total delivered energy (adjusted for charging efficiency and safety margin) not to exceed the usable battery capacity.

Similarly, the fuel budget is
\begin{equation}
\sum_{i \neq j} d_{ij} \,\phi_{\tau(k)} \, x_{ijk}
\;\leq\;
F_{\tau(k)} \,\sigma_{\text{fuel}} \, y_{\tau(k),v(k)}
\quad \forall k \in \mathcal{K},
\end{equation}
which limits the total travel distance to the usable fuel capacity, accounting for reserve margins. Here, $\sigma_{\text{fuel}}$ represents the fuel reserve factor, restricting the usable capacity to 90\% of the nominal tank volume to ensure operational safety margins. The values for $\sigma_{\text{batt}}$ and $\sigma_{\text{fuel}}$ are reported in Table~\ref{tab:mobile_dcfc_specs_corrected}.

\section{Case Study and Scenario Setup}
\label{sec:Solution Approach}
We evaluated the proposed formulation using a case study on the electrification of wheel loader operations at the California Department of Transportation (Caltrans). This section introduces the dataset, scenario construction, and preprocessing steps used for two representative operational regions.

In line with California’s zero-emission mandate (Executive Order N-79-20), mobile fast charging is well suited for heavy off-road equipment such as wheel loaders, whose operations are highly site-constrained and exhibit substantial daily energy demand (typically 100--400+~kWh). Real-world fleet configurations are derived from 2024 Caltrans operational records capturing diesel-powered loader activity, including time-stamped trip segments, geographic positions, and equipment identifiers. Each diesel loader is then mapped to an electric equivalent based on operating weight, engine power, and duty cycle, from which battery capacity and charging requirements are estimated. This translation yields loader-level battery capacities $B_j$ (kWh).

For the case study, battery capacities are converted into off-hour charging demands $E_j$ to reflect a high-demand top-up conducted before or after daily operations. For each client $j$ with battery capacity $B_j$, the required charging amount $E_j$ is defined as
\begin{equation}
E_j = \min\{250,\; \max\{30,\; 0.25\,B_j\}\},
\end{equation}
corresponding to a 25\% battery replenishment, subject to a minimum of 30~kWh and a maximum of 250~kWh to maintain practical service durations during off-hour charging.

Scenario construction follows a unified preprocessing pipeline applied to two contrasting operational contexts: a high-density urban area in Los Angeles and a sparsely distributed mountainous region near Truckee, California. For Los Angeles, loader activity is filtered using a 20-mile radial threshold around Caltrans maintenance facilities and major roadway corridors. Data are aggregated at the daily level, duplicate or conflicting GPS entries are removed, and secondary sites are discarded when loaders appear at distant locations within short time intervals, as such instances are likely attributable to flatbed transport rather than on-site operation.

Two routing scenarios are derived from the processed dataset: (i) a peak-demand case corresponding to the day with the highest loader utilization, and (ii) a statistically representative case, adopted from \cite{abdolmohammadi2024}, in which each loader is assigned to its most frequently observed location. Typical start times, end times, and operation durations are then used to define client-specific off-hour service windows.

Depots are assigned to the largest Caltrans maintenance stations, which serve as operational bases for MFCVs in the routing formulation. For both the Los Angeles and Truckee scenarios, the same filtering and preprocessing framework is applied to maintain methodological consistency.

Several preprocessing procedures are conducted to reduce computational complexity prior to optimization. First, road-network-based travel distances and times between nodes are computed using the OpenRouteService API~\cite{openrouteservice}, assuming symmetric routing to reduce computational burden. Second, compatibility filtering identifies viable vehicle--client pairings based on power capabilities and energy requirements, improving solver initialization without restricting model flexibility. Finally, client service windows are validated to ensure sufficient time for travel, charging, and depot return within the planning horizon, so that only temporally feasible instances are passed to the MILP model.
\begin{figure*}[t]
    \centering
    \includegraphics[width=\textwidth,height=8cm]{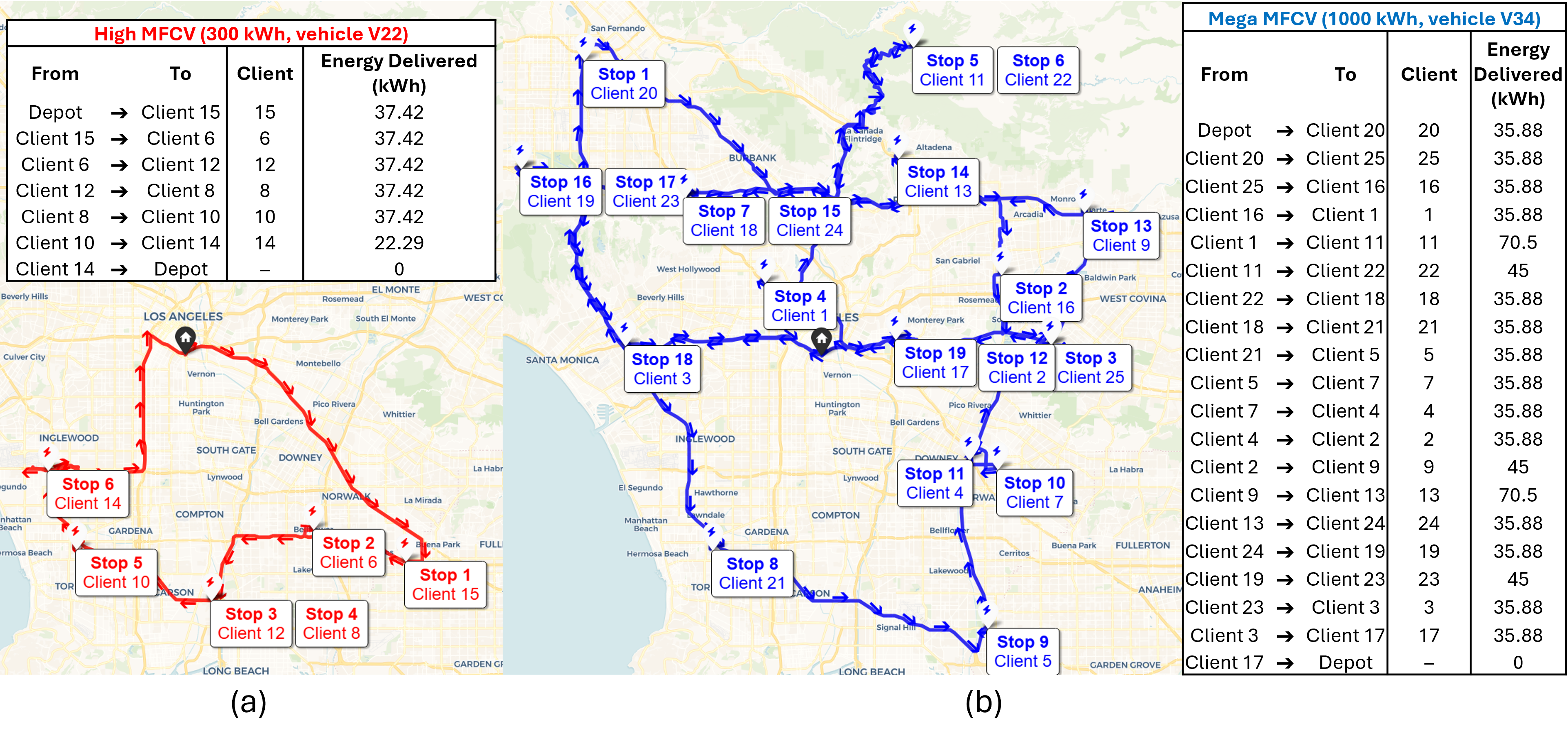}
    \caption{Example illustration of the optimized service routes for the Los Angeles scenario, with arrows showing the direction of travel. Panel (a) shows the optimized route for the High MFCV (300 kWh, vehicle V22), and panel (b) shows the optimized route for the Mega MFCV (1000 kWh, vehicle V34). The corresponding energy delivery details for each route segment are provided in the adjacent tables.}
    \label{fig:result_LA}
\end{figure*}
\subsection*{\textbf{Scenario Characteristics:}}

The two study regions exhibit different spatial and operational patterns that directly shape routing and charging decisions. Los Angeles represents a dense urban environment with clustered demand, moderate inter-site travel distances, and relatively abundant charging opportunities (25 clients within approximately 800~mi$^2$). This concentration facilitates route consolidation and allows multiple clients to be served within a single tour. In contrast, Truckee reflects a rural, mountainous setting with more dispersed demand and longer distances between consecutive stops (6 clients across roughly 210~mi$^2$). These differing profiles yield distinct FSMCVRPTW instances for the two scenarios. Each instance is formulated as a MILP and solved with the Gurobi optimizer via the PuLP Python interface \cite{gurobi}.

\section{Computational Results}
\label{sec:Computational results}

This section evaluated the proposed FSMCVRPTW framework in two contrasting service environments: the high-density Los Angeles region and the rural, geographically dispersed Truckee region. Results demonstrate that the model successfully adapts fleet composition and routing strategies to local demand characteristics. The following subsections examine fleet selection, routing feasibility, temporal performance, and economic outcomes.

\subsection{Fleet Composition Results}
For the Los Angeles scenario, the optimization selects a mixed fleet consisting of one High MFCV (300 kWh) and one Mega MFCV (1000 kWh), as shown in Figure \ref{fig:result_LA}. This configuration achieves full service completion with a 100\% success rate. Table \ref{tab:fleet_composition} reports the selected vehicle types and their utilization, defined as the proportion of mobile DCFC battery capacity used for client charging. Overall routing efficiency and schedule performance are summarized in Table \ref{tab:routing_performance}, which indicates a total travel time of 13.9 hours, 1.38 hours of active service, and 17.1 hours of waiting, with no lateness recorded. Vehicle V22 (High MFCV) served 6 clients and delivered 209 kWh with 69.8\% utilization over 9.97 hours, while Vehicle V34 (Mega MFCV) supplied 778.3 kWh to 19 clients over 22.4 hours, yielding a utilization rate of 77.8\%.

\begin{figure*}[!t]
    \centering
    \begin{subfigure}[t]{0.495\textwidth}
        \centering
        \includegraphics[width=\textwidth, height=6cm]{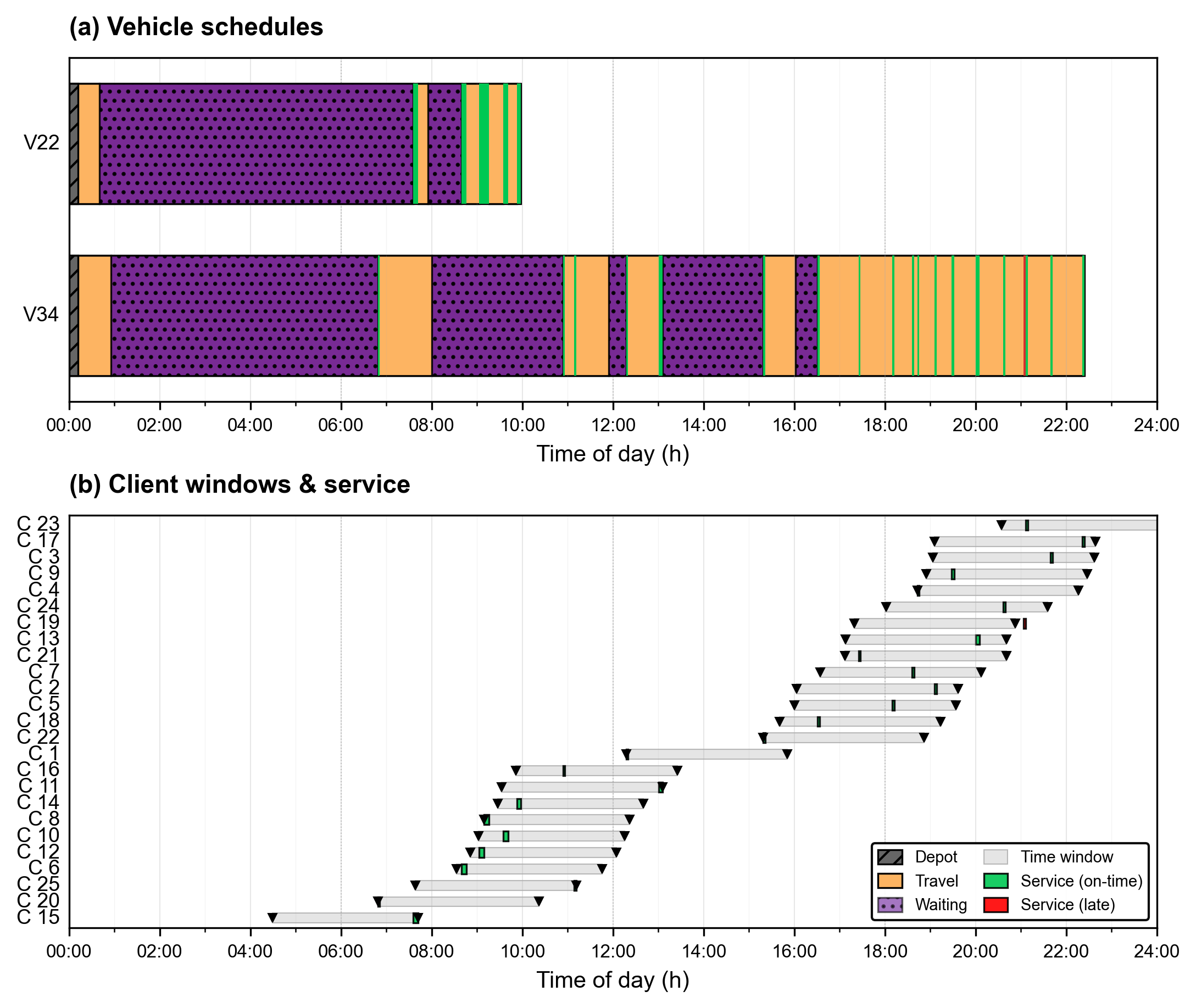}
        \caption{Los Angeles scenario}
        \label{fig:la_timeline}
    \end{subfigure}
    \hfill
    \begin{subfigure}[t]{0.495\textwidth}
        \centering
        \includegraphics[width=\textwidth, height=6cm]{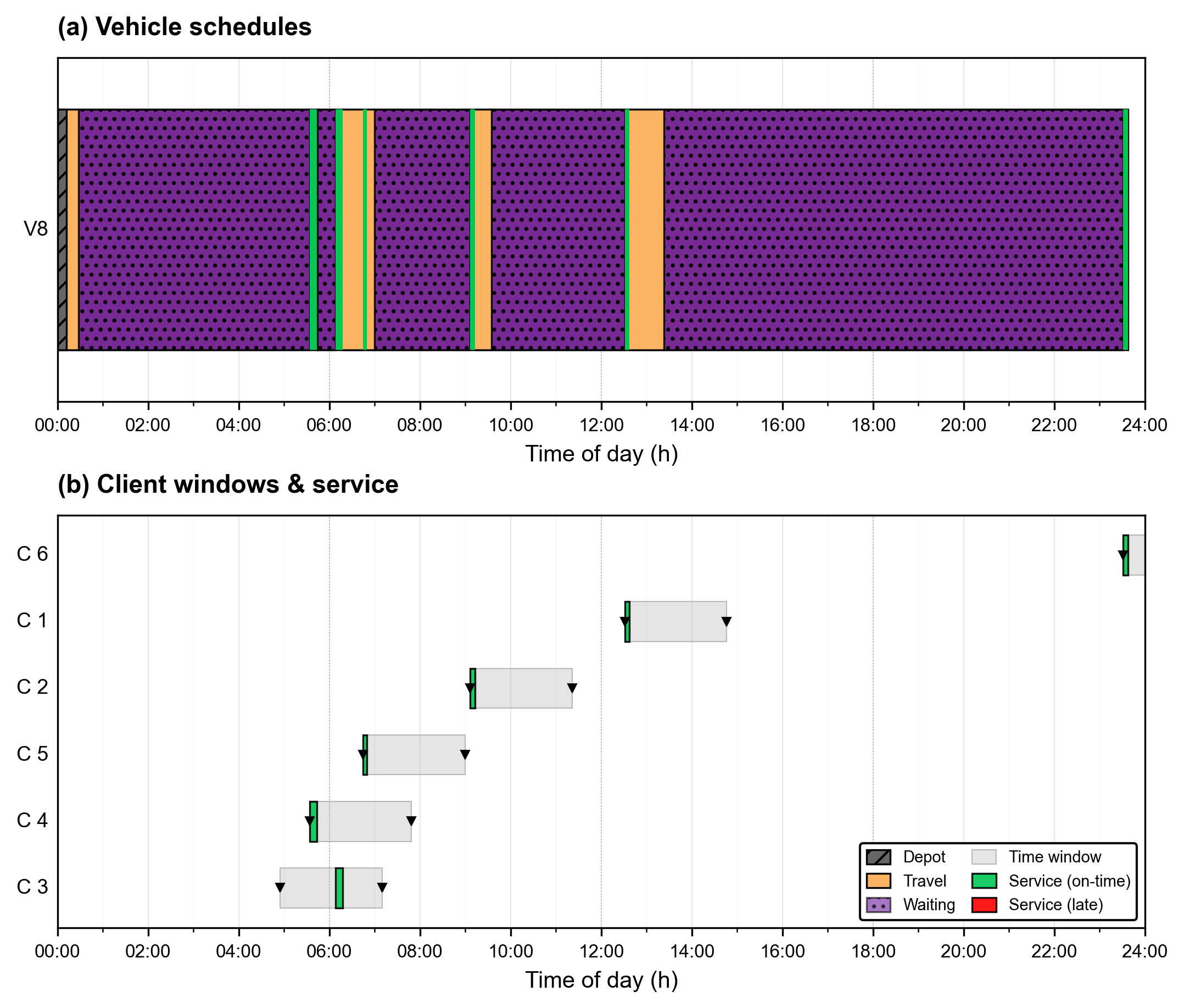}
        \caption{Truckee scenario}
        \label{fig:truckee_timeline}
    \end{subfigure}

    \caption{Comparison of vehicle schedules and client service windows. For both (a) Los Angeles and (b) Truckee, the top panel shows vehicle schedules: depot time (gray hatching), travel (gold), waiting (purple dots), and service (green). The bottom panel compares client time windows (light gray bands) against actual service periods (green bars).}
    \label{fig:timeline_comparison}
\end{figure*}

Figure \ref{fig:la_timeline} illustrates client time windows (light gray) alongside actual charging events (green). All services occurred within assigned windows, confirming full temporal feasibility. However, frequent early arrivals caused by compact routing distances combined with narrow and partially overlapping client time intervals led to extensive waiting, particularly during the early morning for V22 and midday to late afternoon for V34. Consequently, the operational schedule is characterized by short travel, prolonged idle periods, and brief charging durations. This outcome demonstrates the model’s prioritization of temporal compliance over travel-time minimization when operating under strict time-window constraints and highlights the value of a heterogeneous fleet in mitigating temporal bottlenecks.

\begin{figure}
    \centering
    \includegraphics[width=1\linewidth]{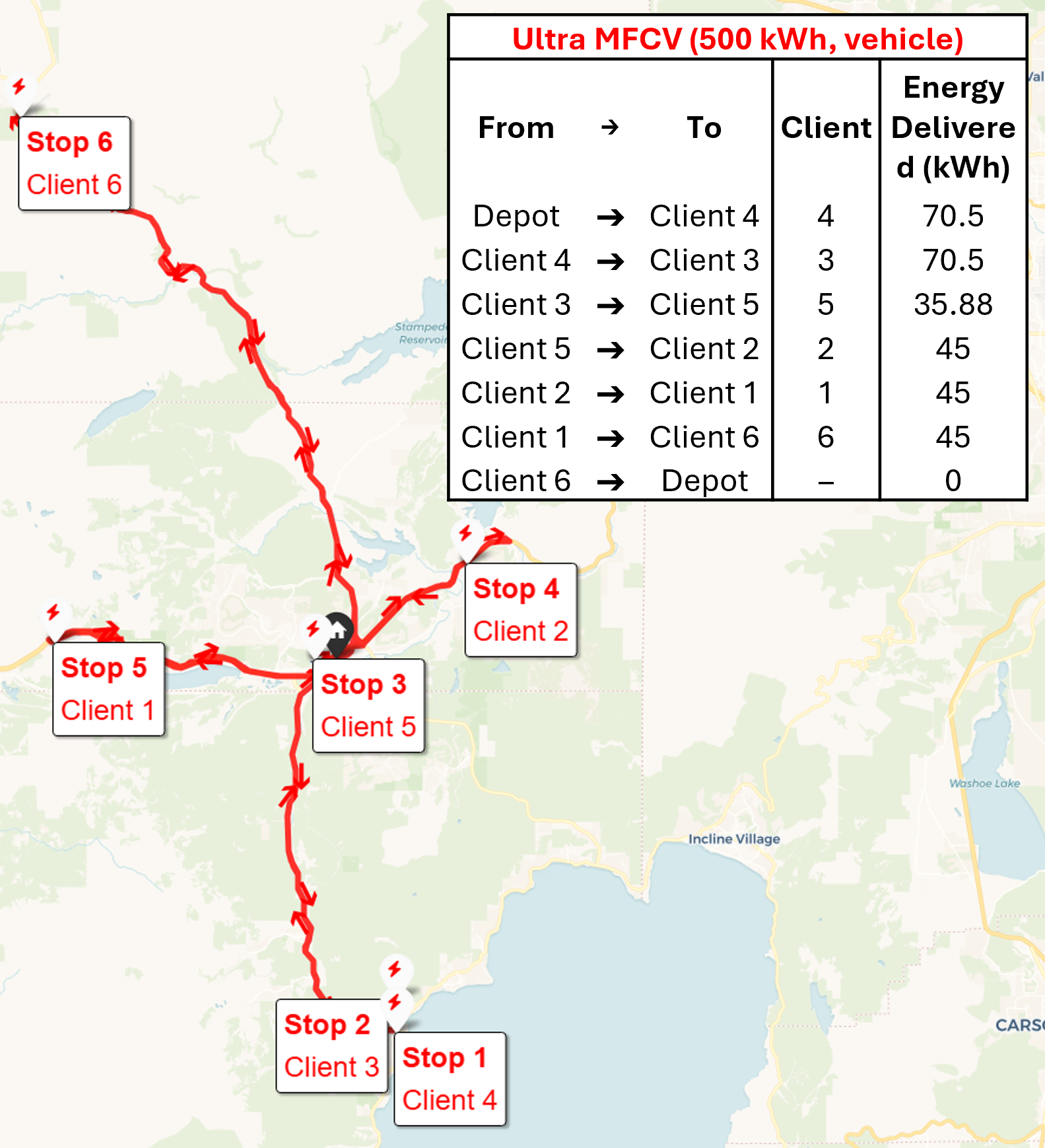}
    \caption{Example illustration of the optimized service route for the Truckee scenario, with arrows indicating the direction of travel. The figure shows the optimized route for the Ultra MFCV (500 kWh), and the corresponding energy delivery information is provided in the embedded table.}
    \label{fig:result_Truckee}
\end{figure}

In contrast, the Truckee scenario features sparse demand and significantly lower aggregate energy requirements. As shown in Figure \ref{fig:result_Truckee}, a single Ultra MFCV (500 kWh) is sufficient to serve all 6 clients with a 100\% completion rate. Table \ref{tab:fleet_composition} confirms that no additional vehicles were required, with the Ultra-class truck achieving 62.3\% utilization and delivering 311.8 kWh across 23.62 hours of operation. Performance metrics in Table \ref{tab:routing_performance} report 4.59 hours of travel, 0.62 hours of service, and 18.41 hours of waiting, again with zero lateness.

Unlike Los Angeles, where short distances lead to early arrival, Truckee’s schedule is dominated by long travel between geographically dispersed clients. Figure \ref{fig:truckee_timeline} shows that waiting primarily results from wide but offset time windows rather than routing density. This operational pattern comprises extended travel followed by relatively short charging periods and limited constraint-driven idle time. Results show the model tailors fleet mix and routes to each region, using one high-capacity vehicle in Truckee and multiple vehicles to handle Los Angeles’ complex time windows.

\begin{table}[!t]
\centering
\caption{Optimal Fleet Composition Results}
\label{tab:fleet_composition}
\begin{tabular}{@{}lcccc@{}}
\toprule
\textbf{Vehicle Type} & \multicolumn{2}{c}{\textbf{Los Angeles}} & \multicolumn{2}{c}{\textbf{Truckee}} \\
\cmidrule(lr){2-3} \cmidrule(lr){4-5}
& \textbf{Count} & \textbf{Utilization} & \textbf{Count} & \textbf{Utilization} \\
\midrule
Standard (80 kWh) & 0 & -- & 0 & -- \\
Medium (160 kWh)  & 0 & -- & 0 & -- \\
High (300 kWh)    & 1 & 69.8\% & 0 & -- \\
Ultra (500 kWh)   & 0 & -- & 1 & 62.3\% \\
Mega (1000 kWh)   & 1 & 77.8\% & 0 & -- \\
\midrule
\textbf{Total fleet size} & \textbf{2} & -- & \textbf{1} & -- \\
\textbf{Amortized CAPEX (/day)} & \textbf{\$928} & -- & \textbf{\$367} & -- \\
\bottomrule
\end{tabular}
\end{table}

\begin{table}[!t]
\centering
\caption{Routing Performance Metrics}
\label{tab:routing_performance}
\begin{tabular}{@{}lcc@{}}
\toprule
\textbf{Metric} & \textbf{Los Angeles} & \textbf{Truckee} \\
\midrule
Total travel time (hours) & 13.9 & 4.59 \\
Total service time (hours) & 1.38 & 0.62 \\
Total waiting time (hours) & 17.1 & 18.41 \\
Service completion rate & 100\% & 100\% \\
Total lateness (hours) & 0 & 0 \\
Fuel consumption (gallons) & 82.7 & 23.13 \\
Energy delivered (kWh) & 988 & 311.88 \\
\bottomrule
\end{tabular}
\end{table}

 \begin{table}[!t]
\centering
\caption{Cost Analysis (\$/day)}
\label{tab:economic_analysis}
\begin{tabular}{@{}lcc@{}}
\toprule
\textbf{Cost Component} & \textbf{Los Angeles} & \textbf{Truckee} \\
\midrule
Travel Time Cost & 417 & 137.69 \\
Service Time Labor & 41.4 & 18.71 \\
Waiting Time Labor & 513.86 & 552.30 \\
Fuel Cost & 314.26 & 87.90 \\
Energy Transfer Cost & 98.8 & 31.19 \\
Lateness Penalties & 0.00 & 0.00 \\
Fleet CAPEX (amortized) & 928.11 & 367.08 \\
Vehicle and Trailer CAPEX (amortized) & 21.92 & 10.96 \\
\midrule
\textbf{Total Daily Cost} & \textbf{2335.35} & \textbf{1205.83} \\
\textbf{Cost per kWh delivered} & \textbf{2.36} & \textbf{3.86} \\
\textbf{Cost per client served} & \textbf{93.41} & \textbf{200.97} \\
\bottomrule
\end{tabular}
\end{table}

\subsection{Economic Analysis}
\label{Economic_Alanysis}
Table \ref{tab:economic_analysis} reports the breakdown of daily operating costs for both scenarios. Los Angeles incurs a substantially higher total daily of 2335.35 \$/day compared with 1205.83 \$/day in Truckee. This difference is primarily driven by the larger fleet requirement in Los Angeles and the significantly higher energy delivery volume. Fleet CAPEX is the dominant cost component, particularly in Los Angeles, where the heterogeneous fleet results in an amortized capital recovery cost of 928.11 \$/day per day, compared with only 367.08 \$/day in Truckee.

Labor-related costs are also non-negligible. Despite benefiting from shorter travel distances, the Los Angeles scenario exhibits similar total waiting labor cost (513.86 \$/day) to Truckee (552.30 \$/day), reflecting the impact of tight and overlapping client time windows in Los Angeles and long inter-site distances in Truckee. Travel time cost is nearly triple in Los Angeles (417 \$/day versus 137.69 \$/day), consistent with the longer cumulative routing duration. Fuel costs are similarly elevated in Los Angeles (314.26 \$/day) due to higher travel frequency and urban driving patterns.

When normalized by service volume, Los Angeles demonstrates superior delivery efficiency. The cost per kWh delivered is 2.36 \$/day compared with 3.86 \$/day in Truckee, and the cost per client served is 93.41 \$/day versus 200.97 \$/day. These values highlight a trade-off between absolute and unit performance: although Truckee achieves lower total operating cost, the sparser demand and longer travel distances lead to a higher cost per unit of energy and per client.

Overall, the Los Angeles configuration leverages economies of scale due to higher service density and energy demand, whereas Truckee requires a more distributed and travel-intensive service strategy, resulting in elevated per-unit delivery costs despite lower total expenditure.

\section{Conclusions and Future Work}
\label{sec: conslusions}
Existing mobile fast charging vehicle (MFCV) planning approaches treat fleet design and operational routing as separate problems, fixing vehicle characteristics before solving dispatch decisions and thereby failing to exploit the coupling between charger specifications and operational strategy.
The fleet size and mix capacitated vehicle routing problem with time windows formulation developed here integrates all decision layers into a single mixed-integer optimization model, solving fleet size, vehicle power capacities, routing sequences, and service schedules simultaneously. The mathematical framework incorporates heterogeneous vehicle types with distinct power ratings, battery capacities, and cost structures while enforcing service window compliance, energy delivery constraints, and network flow conservation within a unified objective function. To illustrate computational tractability and practical applicability of the formulation, we applied it to two case studies constructed from Caltrans operational data for urban and rural deployment contexts.
Results indicate that the unified model scales to realistic problem dimensions while capturing how demand density and spatial structure influence optimal fleet composition. This formulation enables simultaneous minimization of capital expenditure and operational costs within a single optimization, accounting for fleet investment decisions and scheduling, routing, and service time expenses that sequential frameworks must address in isolation. Extensions include stochastic travel time and demand modeling, multi-period planning for charging cycles and maintenance, dynamic routing with real-time information, and grid integration with time-varying electricity prices. The FSMCVRPTW formulation provides a methodological baseline for these extensions and a benchmark for future MFCV planning approaches.

\section*{Acknowledgment}
The authors acknowledge using AI-based tools to enhance manuscript clarity.

\bibliographystyle{IEEEtran}
\bibliography{references}

@misc{IEA:2023,
  author       = {International Energy Agency},
  title        = {{Global EV Outlook 2023}},
  publisher    = {IEA},
  address      = {Paris},
  year         = {2023},
  url          = {https://www.iea.org/reports/global-ev-outlook-2023}
}

@book{IPCC:2022,
  editor       = {Pichs-Madruga, R. and others},
  title        = {{Climate Change 2022: Mitigation of Climate Change. Contribution of Working Group III to the Sixth Assessment Report of the Intergovernmental Panel on Climate Change}},
  publisher    = {Cambridge University Press},
  address      = {Cambridge, UK},
  year         = {2022},
  doi          = {10.1017/9781009157926}
}

@article{Franke:2013,
  author       = {Franke, Thomas and Neumann, Isabel and B{\"u}hler, Franz and Cocron, Paul and Krems, Josef F.},
  title        = {{Experiencing Range in an Electric Vehicle: Understanding Psychological Barriers}},
  journal      = {Applied Psychology: An International Review},
  volume       = {61},
  number       = {3},
  pages        = {368--391},
  year         = {2013},
  doi          = {10.1111/j.1464-0597.2011.00474.x}
}

@techreport{Nicholas:2021,
  author       = {Nicholas, Michael},
  title        = {{Charging Up America: Assessing the Growing Need for U.S. Charging Infrastructure through 2030}},
  institution  = {International Council on Clean Transportation (ICCT)},
  year         = {2021},
  url          = {https://theicct.org/publication/charging-up-america-assessing-the-growing-need-for-u-s-charging-infrastructure-through-2030/}
}

@article{Dong:2014,
  author       = {Dong, Jing and Liu, Changzheng and Lin, Zhenhong},
  title        = {{Charging infrastructure planning for promoting battery electric vehicles: An activity-based approach using multi-day travel data}},
  journal      = {Transportation Research Part C: Emerging Technologies},
  volume       = {38},
  pages        = {44--55},
  year         = {2014},
  doi          = {10.1016/j.trc.2013.11.001}
}

@article{Wang:2020,
  author       = {Wang, H. and Zhao, D. and Meng, Q. and Ong, G. P. and Lee, D. H.},
  title        = {{Network-level energy usage modeling for electric vehicles under various charging scenarios}},
  journal      = {Applied Energy},
  volume       = {275},
  pages        = {115408},
  year         = {2020},
  doi          = {10.1016/j.apenergy.2020.115408}
}

@article{Erdogan:2012,
  author       = {Erdo{\u g}an, Sevgi and Miller-Hooks, Elise},
  title        = {{A Green Vehicle Routing Problem}},
  journal      = {Transportation Research Part E: Logistics and Transportation Review},
  volume       = {48},
  number       = {1},
  pages        = {100--114},
  year         = {2012},
  doi          = {10.1016/j.tre.2011.08.001}
}

@article{Schneider:2014,
  author       = {Schneider, Michael and Stenger, Andreas and Goeke, Dominik},
  title        = {{The Electric Vehicle-Routing Problem with Time Windows and Recharging Stations}},
  journal      = {Transportation Science},
  volume       = {48},
  number       = {4},
  pages        = {500--520},
  year         = {2014},
  doi          = {10.1287/trsc.2013.0490}
}

@article{abdolmohammadi2024,
  title={Sizing and Life Cycle Assessment of Small-Scale Power Backup Solutions: A Statistical Approach},
  author={Abdolmohammadi, Armin and Nemati, Alireza and Haas, Meridian and Nazari, Shima},
  journal={IEEE Access},
  year={2024},
  publisher={IEEE}
}

@book{Toth:2014,
  title={Vehicle Routing: Problems, Methods, and Applications},
  author={Toth, Paolo and Vigo, Daniele},
  year={2014},
  publisher={SIAM},
  edition={2nd},
  address={Philadelphia, PA}
}

@article{Laporte:2009,
  title={Fifty years of vehicle routing},
  author={Laporte, Gilbert},
  journal={Transportation Science},
  volume={43},
  number={4},
  pages={408--416},
  year={2009},
  publisher={INFORMS}
}

@book{Golden:2008,
  title={The Vehicle Routing Problem: Latest Advances and New Challenges},
  author={Golden, Bruce L and Raghavan, S and Wasil, Edward A},
  year={2008},
  publisher={Springer},
  address={Boston, MA}
}

@book{Ahuja:1993,
  title={Network Flows: Theory, Algorithms, and Applications},
  author={Ahuja, Ravindra K and Magnanti, Thomas L and Orlin, James B},
  year={1993},
  publisher={Prentice Hall},
  address={Englewood Cliffs, NJ}
}

@book{Bertsekas:1998,
  title={Network Optimization: Continuous and Discrete Models},
  author={Bertsekas, Dimitri P},
  year={1998},
  publisher={Athena Scientific},
  address={Belmont, MA}
}

@article{Miller:1960,
  title={Integer programming formulation of traveling salesman problems},
  author={Miller, Clair E and Tucker, Albert W and Zemlin, Richard A},
  journal={Journal of the ACM},
  volume={7},
  number={4},
  pages={326--329},
  year={1960},
  publisher={ACM}
}

@article{Desrochers:1991,
  title={Improvements and extensions to the Miller-Tucker-Zemlin subtour elimination constraints},
  author={Desrochers, Martin and Laporte, Gilbert},
  journal={Operations Research Letters},
  volume={10},
  number={1},
  pages={27--36},
  year={1991},
  publisher={Elsevier}
}

@book{Christofides:1981,
  title={The Vehicle Routing Problem},
  author={Christofides, Nicos and Mingozzi, Aristide and Toth, Paolo and Sandi, Claudio},
  year={1981},
  publisher={John Wiley \& Sons},
  address={Chichester, UK}
}

@article{Salhi:2014,
  title={Heuristic algorithms for single and multiple depot vehicle routing problems with pickups and deliveries},
  author={Salhi, Said and Nagy, Gabor},
  journal={European Journal of Operational Research},
  volume={162},
  number={1},
  pages={126--141},
  year={2014},
  publisher={Elsevier}
}

@article{Prins:2004,
  title={A simple and effective evolutionary algorithm for the vehicle routing problem},
  author={Prins, Christian},
  journal={Computers \& Operations Research},
  volume={31},
  number={12},
  pages={1985--2002},
  year={2004},
  publisher={Elsevier}
}

@article{Clarke:1964,
  title={Scheduling of vehicles from a central depot to a number of delivery points},
  author={Clarke, Geoff and Wright, John W},
  journal={Operations Research},
  volume={12},
  number={4},
  pages={568--581},
  year={1964},
  publisher={INFORMS}
}

@article{Fisher:1981,
  title={A generalized assignment heuristic for vehicle routing},
  author={Fisher, Marshall L and Jaikumar, Ramchandran},
  journal={Networks},
  volume={11},
  number={2},
  pages={109--124},
  year={1981},
  publisher={Wiley}
}

@article{Solomon:1987,
  title={Algorithms for the vehicle routing and scheduling problems with time window constraints},
  author={Solomon, Marius M},
  journal={Operations Research},
  volume={35},
  number={2},
  pages={254--265},
  year={1987},
  publisher={INFORMS}
}

@book{Desaulniers:2005,
  title={Column Generation},
  author={Desaulniers, Guy and Desrosiers, Jacques and Solomon, Marius M},
  year={2005},
  publisher={Springer},
  address={New York, NY}
}

@article{Cordeau:2007,
  title={A guide to vehicle routing heuristics},
  author={Cordeau, Jean-Fran{\c{c}}ois and Laporte, Gilbert and Savelsbergh, Martin WP and Vigo, Daniele},
  journal={Journal of the Operational Research Society},
  volume={58},
  number={4},
  pages={512--522},
  year={2007},
  publisher={Taylor \& Francis}
}

@article{Braysy:2005,
  title={Vehicle routing problem with time windows, Part I: Route construction and local search algorithms},
  author={Br{\"a}ysy, Olli and Gendreau, Michel},
  journal={Transportation Science},
  volume={39},
  number={1},
  pages={104--118},
  year={2005},
  publisher={INFORMS}
}

@misc{openrouteservice,
  author       = {{HeiGIT}},
  title        = {OpenRouteService},
  howpublished = {\url{https://openrouteservice.org/}},
  note         = {© openrouteservice.org by HeiGIT. Map data © OpenStreetMap contributors. Licensed under CC-BY 4.0},
  year         = {2025}
}

@article{dantzig1959truck,
  title={The truck dispatching problem},
  author={Dantzig, George B and Ramser, John H},
  journal={Management science},
  volume={6},
  number={1},
  pages={80--91},
  year={1959},
  publisher={Informs}
}

@article{laporte2013vehicle,
  title={Vehicle routing: historical perspective and recent contributions},
  author={Laporte, Gilbert and Toth, Paolo and Vigo, Daniele},
  journal={EURO Journal on Transportation and Logistics},
  volume={2},
  number={1},
  pages={1--4},
  year={2013},
  publisher={Springer}
}

@article{solomon1987algorithms,
  title={Algorithms for the vehicle routing and scheduling problems with time window constraints},
  author={Solomon, Marius M},
  journal={Operations research},
  volume={35},
  number={2},
  pages={254--265},
  year={1987},
  publisher={Informs}
}

@article{tillman1969multiple,
  title={The multiple terminal delivery problem with probabilistic demands},
  author={Tillman, Frank A},
  journal={Transportation Science},
  volume={3},
  number={3},
  pages={192--204},
  year={1969},
  publisher={INFORMS}
}

@article{laporte1984optimal,
  title={Optimal solutions to capacitated multidepot vehicle routing problems},
  author={Laporte, Gilbert},
  journal={Congressus Nemerantium},
  volume={4},
  pages={283--292},
  year={1984}
}

@article{desaulniers2002vrp,
  title={VRP with Pickup and Delivery.},
  author={Desaulniers, Guy and Desrosiers, Jacques and Erdmann, Andreas and Solomon, Marius M and Soumis, Fran{\c{c}}ois},
  journal={The vehicle routing problem},
  volume={9},
  pages={225--242},
  year={2002},
  publisher={Philadelphia}
}

@article{bertsimas1992vehicle,
  title={A vehicle routing problem with stochastic demand},
  author={Bertsimas, Dimitris J},
  journal={Operations Research},
  volume={40},
  number={3},
  pages={574--585},
  year={1992},
  publisher={INFORMS}
}

@book{toth2014vehicle,
  title={Vehicle routing: problems, methods, and applications},
  author={Toth, Paolo and Vigo, Daniele},
  year={2014},
  publisher={SIAM}
}

@article{golden1984fleet,
  title={The fleet size and mix vehicle routing problem},
  author={Golden, Bruce and Assad, Arjang and Levy, Larry and Gheysens, Filip},
  journal={Computers \& Operations Research},
  volume={11},
  number={1},
  pages={49--66},
  year={1984},
  publisher={Elsevier}
}

@article{arabani2012facility,
  title={Facility location dynamics: An overview of classifications and applications},
  author={Arabani, Alireza Boloori and Farahani, Reza Zanjirani},
  journal={Computers \& Industrial Engineering},
  volume={62},
  number={1},
  pages={408--420},
  year={2012},
  publisher={Elsevier}
}

@article{kucukoglu2021electric,
  title={The electric vehicle routing problem and its variations: A literature review},
  author={Kucukoglu, Ilker and Dewil, Reginald and Cattrysse, Dirk},
  journal={Computers \& Industrial Engineering},
  volume={161},
  pages={107650},
  year={2021},
  publisher={Elsevier}
}

@article{erdougan2012green,
  title={A green vehicle routing problem},
  author={Erdo{\u{g}}an, Sevgi and Miller-Hooks, Elise},
  journal={Transportation research part E: logistics and transportation review},
  volume={48},
  number={1},
  pages={100--114},
  year={2012},
  publisher={Elsevier}
}

@article{schneider2014electric,
  title={The electric vehicle-routing problem with time windows and recharging stations},
  author={Schneider, Michael and Stenger, Andreas and Goeke, Dominik},
  journal={Transportation science},
  volume={48},
  number={4},
  pages={500--520},
  year={2014},
  publisher={INFORMS}
}

@article{felipe2014heuristic,
  title={A heuristic approach for the green vehicle routing problem with multiple technologies and partial recharges},
  author={Felipe, {\'A}ngel and Ortu{\~n}o, M Teresa and Righini, Giovanni and Tirado, Gregorio},
  journal={Transportation Research Part E: Logistics and Transportation Review},
  volume={71},
  pages={111--128},
  year={2014},
  publisher={Elsevier}
}

@article{montoya2017electric,
  title={The electric vehicle routing problem with nonlinear charging function},
  author={Montoya, Alejandro and Gu{\'e}ret, Christelle and Mendoza, Jorge E and Villegas, Juan G},
  journal={Transportation Research Part B: Methodological},
  volume={103},
  pages={87--110},
  year={2017},
  publisher={Elsevier}
}

@article{keskin2016partial,
  title={Partial recharge strategies for the electric vehicle routing problem with time windows},
  author={Keskin, Merve and {\c{C}}atay, B{\"u}lent},
  journal={Transportation research part C: emerging technologies},
  volume={65},
  pages={111--127},
  year={2016},
  publisher={Elsevier}
}

@article{goeke2015routing,
  title={Routing a mixed fleet of electric and conventional vehicles},
  author={Goeke, Dominik and Schneider, Michael},
  journal={European Journal of Operational Research},
  volume={245},
  number={1},
  pages={81--99},
  year={2015},
  publisher={Elsevier}
}

@article{hiermann2016electric,
  title={The electric fleet size and mix vehicle routing problem with time windows and recharging stations},
  author={Hiermann, Gerhard and Puchinger, Jakob and Ropke, Stefan and Hartl, Richard F},
  journal={European Journal of Operational Research},
  volume={252},
  number={3},
  pages={995--1018},
  year={2016},
  publisher={Elsevier}
}

@article{cui2018mobile,
  title={The mobile charging vehicle routing problem with time windows and recharging services},
  author={Cui, Shaohua and Zhao, Hui and Chen, Hui and Zhang, Cuiping},
  journal={Computational intelligence and neuroscience},
  volume={2018},
  number={1},
  pages={5075916},
  year={2018},
  publisher={Wiley Online Library}
}

@article{afshar2021mobile,
  title={Mobile charging stations for electric vehicles—A review},
  author={Afshar, Shahab and Macedo, Pablo and Mohamed, Farog and Disfani, Vahid},
  journal={Renewable and Sustainable Energy Reviews},
  volume={152},
  pages={111654},
  year={2021},
  publisher={Elsevier}
}

@article{huang2014design,
  title={Design of a mobile charging service for electric vehicles in an urban environment},
  author={Huang, Shisheng and He, Liang and Gu, Yu and Wood, Kristin and Benjaafar, Saif},
  journal={IEEE Transactions on Intelligent Transportation Systems},
  volume={16},
  number={2},
  pages={787--798},
  year={2014},
  publisher={IEEE}
}

@article{ruaboacua2020optimization,
  title={An optimization model for the temporary locations of mobile charging stations},
  author={R{\u{a}}boac{\u{a}}, Maria-Simona and B{\u{a}}ncescu, Irina and Preda, Vasile and Bizon, Nicu},
  journal={Mathematics},
  volume={8},
  number={3},
  pages={453},
  year={2020},
  publisher={MDPI}
}

@inproceedings{afshar2022optimal,
  title={Optimal scheduling of electric vehicles in the presence of mobile charging stations},
  author={Afshar, Shahab and Disfani, Vahid},
  booktitle={2022 IEEE Power \& Energy Society General Meeting (PESGM)},
  pages={1--5},
  year={2022},
  organization={IEEE}
}

@article{tang2020online,
  title={Online-to-offline mobile charging system for electric vehicles: Strategic planning and online operation},
  author={Tang, Peng and He, Fang and Lin, Xi and Li, Meng},
  journal={Transportation Research Part D: Transport and Environment},
  volume={87},
  pages={102522},
  year={2020},
  publisher={Elsevier}
}

@article{beyazit2023electric,
  title={Electric vehicle charging through mobile charging station deployment in coupled distribution and transportation networks},
  author={Beyaz{\i}t, Muhammed Ali and Ta{\c{s}}c{\i}karao{\u{g}}lu, Ak{\i}n},
  journal={Sustainable Energy, Grids and Networks},
  volume={35},
  pages={101102},
  year={2023},
  publisher={Elsevier}
}

@article{liu2024multi,
  title={Multi-agent deep reinforcement learning based scheduling approach for mobile charging in internet of electric vehicles},
  author={Liu, Linfeng and Huang, Zhuo and Xu, Jia},
  journal={IEEE Transactions on Mobile Computing},
  volume={23},
  number={10},
  pages={10130--10145},
  year={2024},
  publisher={IEEE}
}

@article{xu2025dynamic,
  title={Dynamic Expansion Planning of Charging Stations With Fixed and Mobile Chargers},
  author={Xu, Jiuping and Ma, Jianglin and Zhao, Chuandang and Li, Yuhan},
  journal={IEEE Transactions on Intelligent Transportation Systems},
  year={2025},
  publisher={IEEE}
}

@article{duan2025study,
  title={A study on mobile charging station combined with integrated energy system: Emphasis on energy dispatch strategy and multi-scenario analysis},
  author={Duan, Sudong and Zhang, Zhonghui and Wang, Zhaojun and Xiong, Xiaoyue and Chen, Xinhan and Que, Xiaoyu},
  journal={Renewable Energy},
  volume={239},
  pages={122111},
  year={2025},
  publisher={Elsevier}
}

@article{pelletier2019electric,
  title={The electric vehicle routing problem with energy consumption uncertainty},
  author={Pelletier, Samuel and Jabali, Ola and Laporte, Gilbert},
  journal={Transportation Research Part B: Methodological},
  volume={126},
  pages={225--255},
  year={2019},
  publisher={Elsevier}
}

@article{schiffer2017electric,
  title={The electric location routing problem with time windows and partial recharging},
  author={Schiffer, Maximilian and Walther, Grit},
  journal={European journal of operational research},
  volume={260},
  number={3},
  pages={995--1013},
  year={2017},
  publisher={Elsevier}
}

@article{basso2019energy,
  title={Energy consumption estimation integrated into the electric vehicle routing problem},
  author={Basso, Rafael and Kulcs{\'a}r, Bal{\'a}zs and Egardt, Bo and Lindroth, Peter and Sanchez-Diaz, Ivan},
  journal={Transportation Research Part D: Transport and Environment},
  volume={69},
  pages={141--167},
  year={2019},
  publisher={Elsevier}
}

@article{verma2018electric,
  title={Electric vehicle routing problem with time windows, recharging stations and battery swapping stations},
  author={Verma, Amit},
  journal={EURO Journal on Transportation and Logistics},
  volume={7},
  number={4},
  pages={415--451},
  year={2018},
  publisher={Springer}
}

@article{zhang2020fuzzy,
  title={Fuzzy optimization model for electric vehicle routing problem with time windows and recharging stations},
  author={Zhang, Shuai and Chen, Mingzhou and Zhang, Wenyu and Zhuang, Xiaoyu},
  journal={Expert systems with applications},
  volume={145},
  pages={113123},
  year={2020},
  publisher={Elsevier}
}

@ARTICLE{Schneider2018-zp,
  title     = "Optimization of battery charging and purchasing at electric
               vehicle battery swap stations",
  author    = "Schneider, Frank and Thonemann, Ulrich W and Klabjan, Diego",
  journal   = "Transp. Sci.",
  publisher = "Institute for Operations Research and the Management Sciences
               (INFORMS)",
  volume    =  52,
  number    =  5,
  pages     = "1211--1234",
  month     =  oct,
  year      =  2018,
  language  = "en"
}

@article{fisher1981generalized,
  title={A generalized assignment heuristic for vehicle routing},
  author={Fisher, Marshall L and Jaikumar, Ramchandran},
  journal={Networks},
  volume={11},
  number={2},
  pages={109--124},
  year={1981},
  publisher={Wiley Online Library}
}

@article{salhi2013fleet,
  title={The fleet size and mix vehicle routing problem with backhauls: formulation and set partitioning-based heuristics},
  author={Salhi, Said and Wassan, Niaz and Hajarat, Mutaz},
  journal={Transportation Research Part E: Logistics and Transportation Review},
  volume={56},
  pages={22--35},
  year={2013},
  publisher={Elsevier}
}

@article{zhang2025joint,
  title={Joint optimization of fleet sizing, charging station planning, and operation for autonomous electric vehicle fleets in urban transportation networks},
  author={Zhang, Huayu and Jin, Ding and Han, Bing and Xue, Fei and Lu, Shaofeng and Jiang, Lin},
  journal={Sustainable Energy, Grids and Networks},
  pages={101946},
  year={2025},
  publisher={Elsevier}
}

@article{yang2023fleet,
  title={Fleet sizing and charging infrastructure design for electric autonomous mobility-on-demand systems with endogenous congestion and limited link space},
  author={Yang, Jie and Levin, Michael W and Hu, Lu and Li, Haobin and Jiang, Yangsheng},
  journal={Transportation Research Part C: Emerging Technologies},
  volume={152},
  pages={104172},
  year={2023},
  publisher={Elsevier}
}

@article{Schiffer:2018,
  author    = {Michael Schiffer and G{"u}nther W. Weber and Thomas C. Schlechte},
  title     = {A Survey on Electric Vehicle Fleet Management for Smart Cities and Autonomous Vehicles},
  journal   = {European Journal of Operational Research},
  year      = {2018},
  volume    = {274},
  number    = {2},
  pages     = {491--507},
  doi       = {10.1016/j.ejor.2018.05.043}
}

@misc{gurobi,
  author = {{Gurobi Optimization, LLC}},
  title = {{Gurobi Optimizer Reference Manual}},
  year = 2024,
  url = "https://www.gurobi.com"
}

@article{Yousefgomgashte,
title = "PACE: A Framework for Learning and Control in Linear Incomplete-Information Differential Games",
year = "2025",
language = "English (US)",
volume = "283",
pages = "1419--1433",
journal = "Proceedings of Machine Learning Research",
issn = "2640-3498",
publisher = "ML Research Press",
}

\end{document}